%% file: imt.tex
\documentclass[sigconf]{acmart}

\setcopyright{none}
\renewcommand\footnotetextcopyrightpermission[1]{} % removes footnote with conference information in first column
\usepackage{balance}
\usepackage{booktabs} % For formal tables
\usepackage{listings}
\usepackage{color}
\usepackage[skip=2pt]{caption}
\usepackage{subcaption}
\usepackage{cleveref}
\usepackage{textcomp}
\usepackage{graphicx}

\usepackage[ruled, lined, linesnumbered, commentsnumbered, longend]{algorithm2e}
\usepackage{xcolor}
\newlength\mylen
\newcommand\myinput[1]{%
   \settowidth\mylen{\KwIn{}}%
   \setlength\hangindent{\mylen}%
   \hspace*{\mylen}#1\\}
\newcommand\myoutput[1]{%
   \settowidth\mylen{\KwOut{}}%
   \setlength\hangindent{\mylen}%
   \hspace*{\mylen}#1\\}
   
\usepackage{booktabs,siunitx,array,threeparttable}
\usepackage{paralist}
\usepackage{amsthm}
\newtheorem{property}{Property}

\def\BibTeX{{\rm B\kern-.05em{\sc i\kern-.025em b}\kern-.08emT\kern-.1667em\lower.7ex\hbox{E}\kern-.125emX}}
\DeclareCaptionFormat{captionformat}{\fontsize{7}{8}\selectfont#1#2#3}
\definecolor{dkgreen}{rgb}{0,0.6,0}
\definecolor{gray}{rgb}{0.5,0.5,0.5}
\definecolor{mauve}{rgb}{0.58,0,0.82}

\usepackage{enumitem}
\setlist{topsep=0pt, leftmargin=15pt}

\begin{document}
\linepenalty=100
\setlength{\abovedisplayskip}{3pt}
\setlength{\belowdisplayskip}{3pt}

\title{Learning to Rank for Maps at Airbnb}

\author{Malay Haldar, Hongwei Zhang, Kedar Bellare, Sherry Chen, Soumyadip Banerjee, Xiaotang Wang,
  Mustafa Abdool, Huiji Gao, Pavan Tapadia, Liwei He, Sanjeev Katariya}
\affiliation{%
  \institution{Airbnb, Inc.}
  \city{San Francisco}
  \state{CA}
    \country{USA}
}
\email{malay.haldar@airbnb.com}

\renewcommand{\shortauthors}{Malay Haldar et al.}

\begin{abstract}
As a two-sided marketplace, Airbnb brings together hosts who own listings for rent with prospective guests from around the globe. Results from a guest’s search for listings are displayed primarily through two interfaces: (1) as a list of rectangular cards that contain on them the listing image, price, rating, and other details, referred to as {\it list-results} (2) as oval pins on a map showing the listing price, called {\it map-results}. Both these interfaces, since their inception, have used the same ranking algorithm that orders listings by their booking probabilities and selects the top listings for display. But some of the basic assumptions underlying ranking, built for a world where search results are presented as lists, simply break down for maps. This paper describes how we rebuilt ranking for maps by revising the mathematical foundations of how users interact with search results. Our iterative and experiment-driven approach led us through a path full of twists and turns, ending in a unified theory for the two interfaces. Our journey shows how assumptions taken for granted when designing machine learning algorithms may not apply equally across all user interfaces, and how they can be adapted. The net impact was one of the largest improvements in user experience for Airbnb which we discuss as a series of experimental validations.
\end{abstract}




\maketitle
\pagestyle{plain}
\pagenumbering{gobble}

\input{imtbody-conf}

%\bibliographystyle{ACM-Reference-Format}
\bibliographystyle{ACM-Reference-Format}

\balance 
\bibliography{imt-bibliography}

\end{document}

%% file: imtbody-conf.tex
\section{Introduction}
As of June 01, 2024, there are more than $7.7$ million active Airbnb listings in over $100$ thousand cities and towns worldwide. The total number of guest arrivals, contributed by guests all across the world, now exceeds $1.5$ billion. The vast majority of these guest arrivals started with a search, the core mechanism that connects guests to hosts. What makes Airbnb search a particularly strong case for machine learning application is its global scale, coupled with the individuality of each listing. Every listing has its unique location, along with its own look and feel. Furthermore, even the most popular listing can only be booked a maximum of $365$ days in a year, so memorizing top results and replaying them back is not an option. Airbnb search demands a truly generalized learning of what each listing is offering.

And true to expectations, artificial intelligence shines on this occasion, far surpassing what human intelligence can achieve. As part of an evaluation exercise, ranking engineers were shown pairs of listings and asked to identify which listing out of the pair was booked by a searcher. The ranking model correctly identified the booked listing $88\%$ of the time, while ranking engineers could manage only $70\%$.

This level of performance wasn’t achieved overnight.  The launch of our first neural network model is described in \cite{kdd19}, and its evolution is captured by ~\cite{kdd20}, ~\cite{kdddiversity}, ~\cite{kdd23}, and ~\cite{cikmdiversity}.

\begin{figure}
    %%\centering
       \begin{minipage}{2in}
       \rule{0pt}{1em}\vspace{-5.2em}
       \begin{flushleft}
        \centering
       \includegraphics[height=0.3in, width=2in]{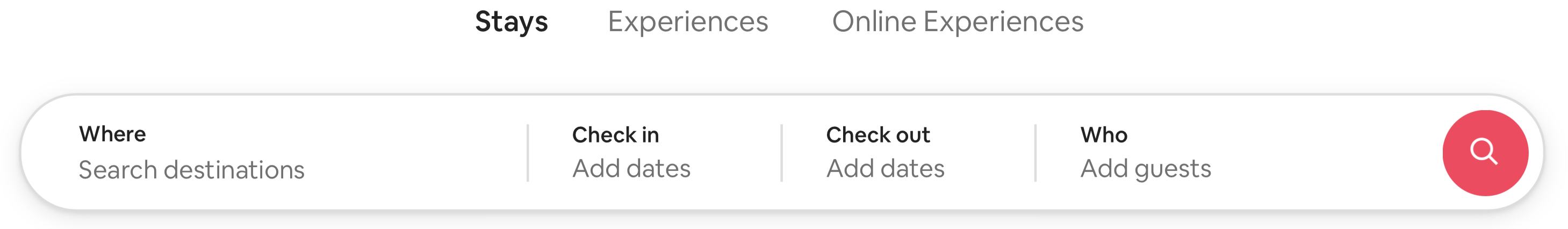}
       \caption{\textmd{Search box with destination, checkin/checkout dates and guest count as inputs.}}
       \label{fig:searchbox}

       \rule{0pt}{1em}\vspace{1em}
       \includegraphics[height=2in, width=2in]{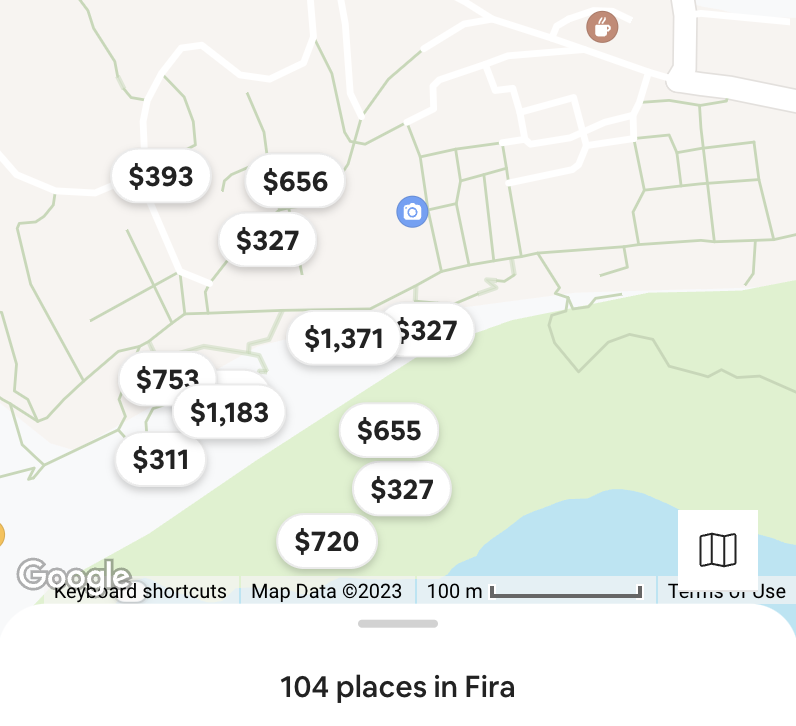}
        \caption{\textmd{Search results as map pins.}}
        \label{fig:map-result}
        \end{flushleft}
    \end{minipage}
     \begin{minipage}{1.3in}
        \begin{flushright}
        \centering
        \includegraphics[height=3.2in, width=1.3in]{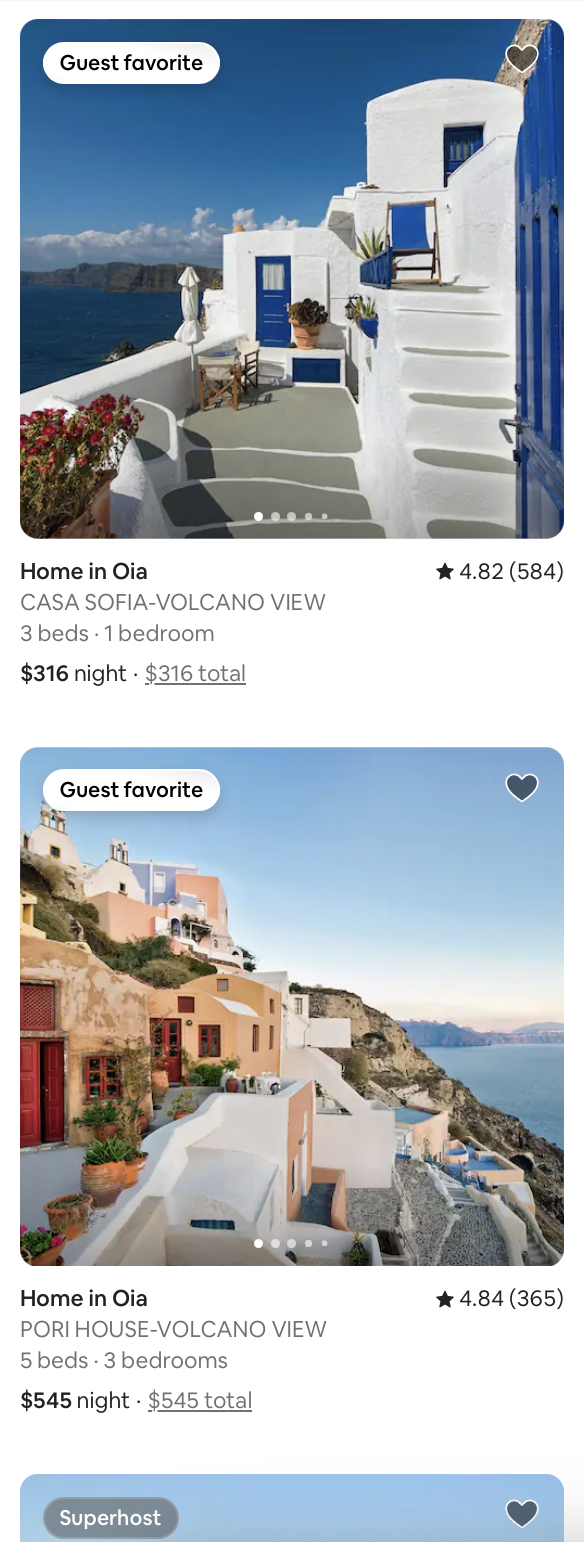} 
        \caption{\textmd{Search results as list of cards.}}
        \label{fig:list-result}
        \end{flushright}
    \end{minipage}
\end{figure}

The ranking tech stack described in these publications powers searches from two sources. First is the familiar search box, where a destination location, check-in/checkout dates, and guest counts are entered explicitly by the searcher (Figure~\ref{fig:searchbox}). Results from the search can be accessed as list-results (Figure~\ref{fig:list-result}), and map-results (Figure~\ref{fig:map-result}). The user can then point the map to an area, and the latitude and longitude boundaries of the map serve as a second source of searches with implicit query parameters. Overall the search box generates $20\%$ of searches, the rest coming from maps.

The main goal of these interfaces is to maximize the number of bookings---the key measure of success both from the guest and the host perspective. To achieve this goal, listings are sorted by their booking probabilities, and the top listings are selected for display as cards in a list and as map pins. The algorithm works by ensuring that the higher the booking probability of a listing, the more attention it receives from users. A formal analysis is presented in Section 3 of ~\cite{cikmdiversity}. This algorithm remained the status quo for many years. But as it happens, disruptions come unannounced.

\section{Map $\neq$ List} \label{mapnelist}
Normalized discounted cumulative gain (NDCG) is our de facto tool for evaluating ranking. To understand ranking performance in various settings, we look at NDCG for that query segment. Comparing the NDCG for queries from the search box against the NDCG for map-generated queries uncovers a puzzle---a stubborn gap of $2\%$.

For many years, engineers suspected the tech stack powering search through maps ought to include something unique to the interface itself. But this mostly drove feature engineering focused on maps, such as distance of the listing from the map center. The general wisdom was that as long as all the necessary features were supplied to the model, it would “do the right thing.” There wasn’t much to explore beyond that.

Although map search interfaces are used by billions every day, there is little mention of how to customize search for maps in the machine learning literature. To the best of our knowledge, there are some references from the early days of the internet, like ~\cite{yates2000searching}, but there seems to be no research trail for the last twenty years. Discussion of ranking for products similar to Airbnb (\cite{mavridis2020beyond}, ~\cite{bernardi2019150}, ~\cite{ursu2018power}) give no special consideration to maps. This lack of prior publications further supports the thinking that map search is nothing special. And yet, we have hints that searching using maps leads to a different user behavior.

\begin{figure}
\includegraphics[height=1.5in, width=3in]{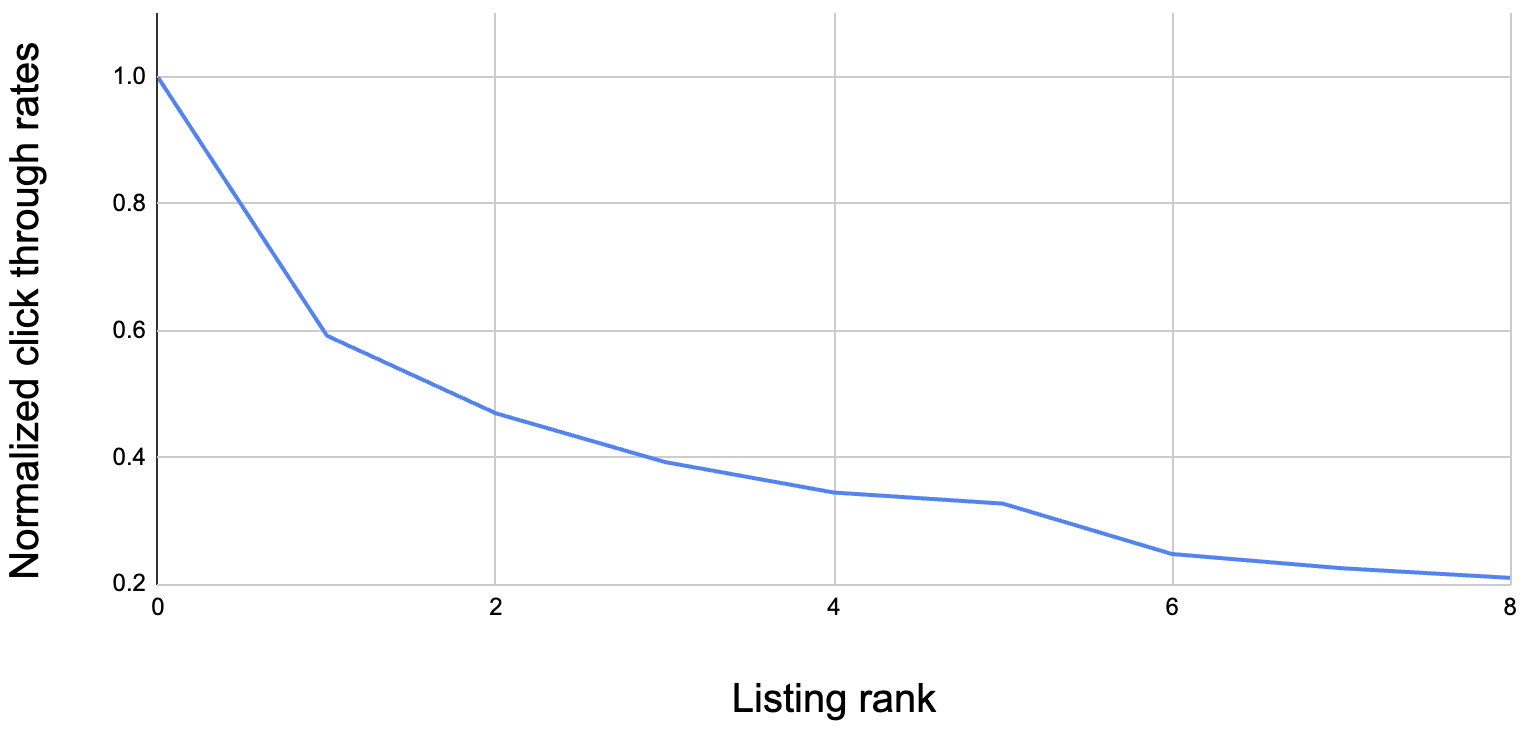}
\caption{\textmd{X-axis: Search rank for list-results. Y-axis: Normalized click-through rates.}}
\label{fig:ctrdiscountcurve}
\end{figure}

To shift tactics, we move away from trying to equate NDCG across the two interfaces, and instead ask whether we should even fixate on the NDCG for map-results? For list-results, our observations are in line with those reported in the past. A plot of click-through rate per ranking position in Figure~\ref{fig:ctrdiscountcurve} visualizes the decay of user attention for list-results, in agreement with established research findings (\cite{clickmodel1}, ~\cite{papoutsaki2017searchgazer}). Section 3 in ~\cite{cikmdiversity} describes how this monotonic decay of user attention is key to the workings of NDCG.

But what about map-results, where the decay of attention is no longer applicable? Since user attention decay is at the heart of NDCG, the metric is no longer meaningful in the context of maps. In fact, {\it ranking itself is irrelevant for map-results,} as there is no sequential list involved.

To be clear, ranking still plays the crucial role of selecting the top listings for display from all the listings available in the map area. But once the top listings are selected, their relative ordering is irrelevant. In the next section, we experimentally validate the hypothesis that ranking doesn’t matter for map-results.

\subsection{Experimental Results}\label{mapnelistexpr}
The validation is a simple online A/B experiment where the treatment randomly shuffles the top results for searches originating from the map. The experiment is restricted to mobile platforms like iOS\texttrademark ~and Android\texttrademark ~where users cannot interact with both list-results and map-results at the same time. The case of desktop web browsers, where concurrent access to list-results and map-results is possible, will be discussed later in Section~\ref{minipin}. From past experiments, we know such a randomization applied to queries from the search box can produce a jaw-dropping booking loss of ~8\% and degrade NDCG by as much as ~5\%. The hypothesis is that for map searches, randomization will produce no difference.

And this experiment indeed confirms the hypothesis. Metrics in treatment are almost identical to control, and more specifically, there is no difference in bookings across treatment and control. This firmly establishes that for searches through the map, ranking the top listings has no effect at all.

\begin{figure}
\includegraphics[height=1.5in, width=3in]{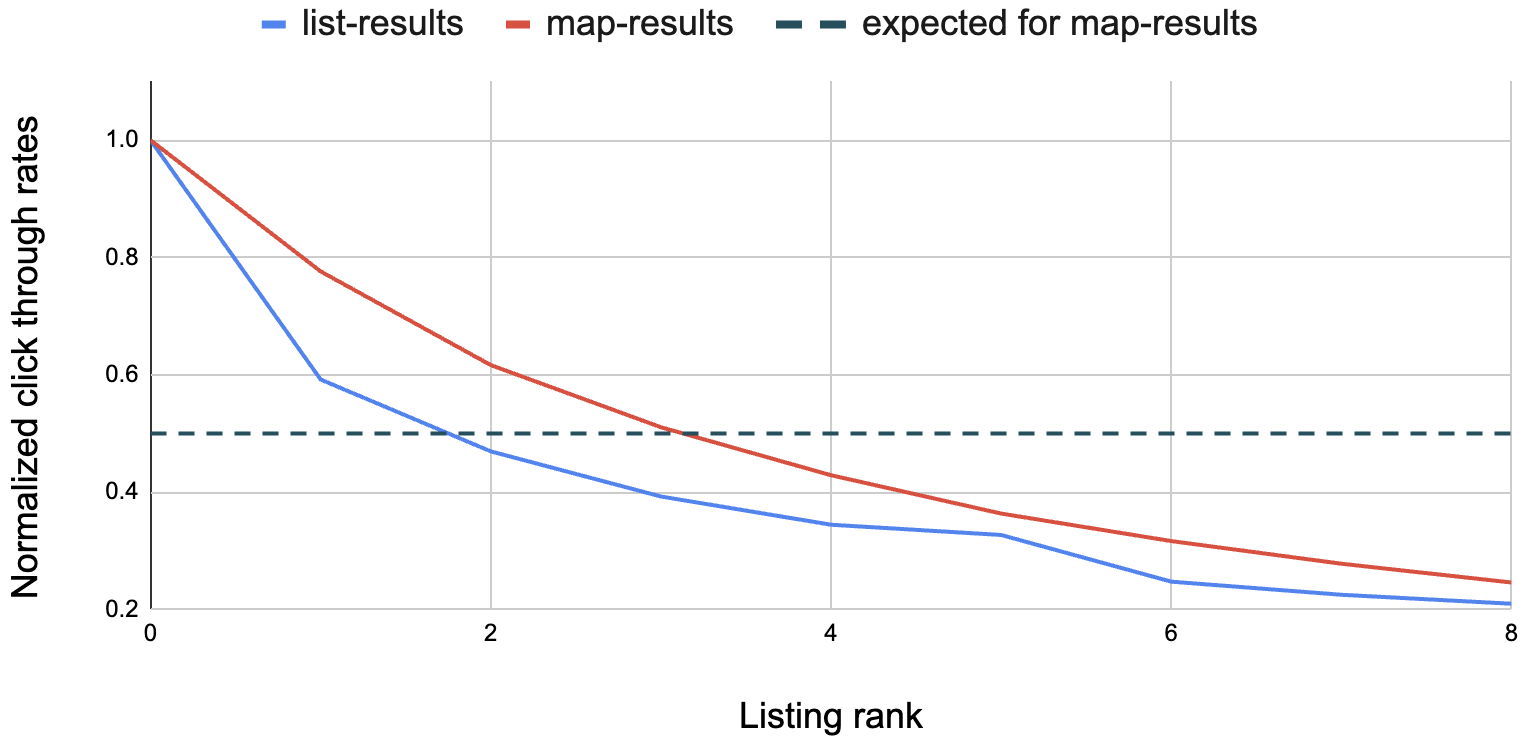}
\caption{\textmd{X-axis: Search rank for list-results and map-results. Y-axis: Normalized click-through rates.}}
\label{fig:mapvslistcurve}
\end{figure}

To show that ranking is irrelevant for map pins, an alternative to the randomization experiment is to plot a rank vs. CTR plot for map-results, along the lines of Figure~\ref{fig:ctrdiscountcurve}. For map-results, we expect the listing ranks to have no connection to user attention, and hence the CTR for each position to be the same, making the plot a straight horizontal line. Instead we get Figure~\ref{fig:mapvslistcurve}! The CTR by ranking position for map-results continues to slope downwards.

We resolve the mystery of this seeming contradiction in Section~\ref{mapattn}. For now, let's take the hypothesis we proved through the randomization experiment and put it into action.

\section{Less Is More}\label{lim}
Consider a very simplified view of the probability of booking a listing via a given query $Q$. All listings eligible for the query are sorted by their booking probabilities and the top $N$ selected for display. We use the following notation:\newline
$\{l_1, l_2, …, l_{N}\} $: Listings ranked $1$ through $N$. \newline
$P_{attention}(i)$ : The relative attention received at rank $i$. \newline
$P_{booking}(l_i)$ : The probability of booking for $l_i$. \newline

We write the probability of booking a listing via $Q$ using an approach similar to Section 5.1 of ~\cite{ltrdebias} and Section 3 of ~\cite{cikmdiversity}:
\begin{equation}\label{eq1}
P_{booking}(Q) = \sum \limits_{i=1}^{N} P_{attention}(i) * P_{booking}(l_i)
\end{equation}

For list-results, $P_{attention}(i) > P_{attention}(j) ~\forall~ i < j$ (\cite{clickmodel2}, ~\cite{lorigo2008eye}). Ordering the listings by $P_{booking}(l_i)$ maximizes $P_{booking}(Q)$, as it iteratively matches the largest probability of booking with the largest user attention. Disrupting this ordering lowers $P_{booking}(Q)$, and hence lowers the observed bookings when tested online.

If $P_{attention}(i) > P_{attention}(j) ~\forall~ i < j$ was true for map-results as well, matching of $P_{booking}(l_i)$ with $P_{attention}(i)$ would be suboptimal in the randomization experiment, causing $P_{booking}(Q)$ to drop, resulting in a bookings loss for the treatment. But since bookings didn't drop in the randomization experiment, the property $P_{attention}(i) > P_{attention}(j) ~\forall~ i < j$ must be false for map-results. Instead, the following must hold $P_{attention}(i) = P_{attention}(j) ~\forall~ i, j  \in \{1, \dots, N\}$. To represent that user attention is distributed equally across the top listings displayed as map pins, we assign $P_{attention}(i) = 1/N$, which transforms Equation~\ref{eq1} for maps to:

\begin{equation}\label{eq2}
P_{booking}(Q) = \frac{1}{N} * \sum \limits_{i=1}^{N} P_{booking}(l_i)
\end{equation}

\begin{property}\label{prop1}
The probability of booking a listing from a map-result is given by the average booking probability of the map pins.
\end{property}

Given that listings with the highest booking probabilities are selected as map pins, and because of Property~\ref{prop1}, the lower the number of pins, the higher the average booking probability. This leads to the {\it Less Is More} principle of map-results: the lesser the number of pins shown, the more the bookings.

\begin{figure}
\includegraphics[height=1.5in, width=2.75in]{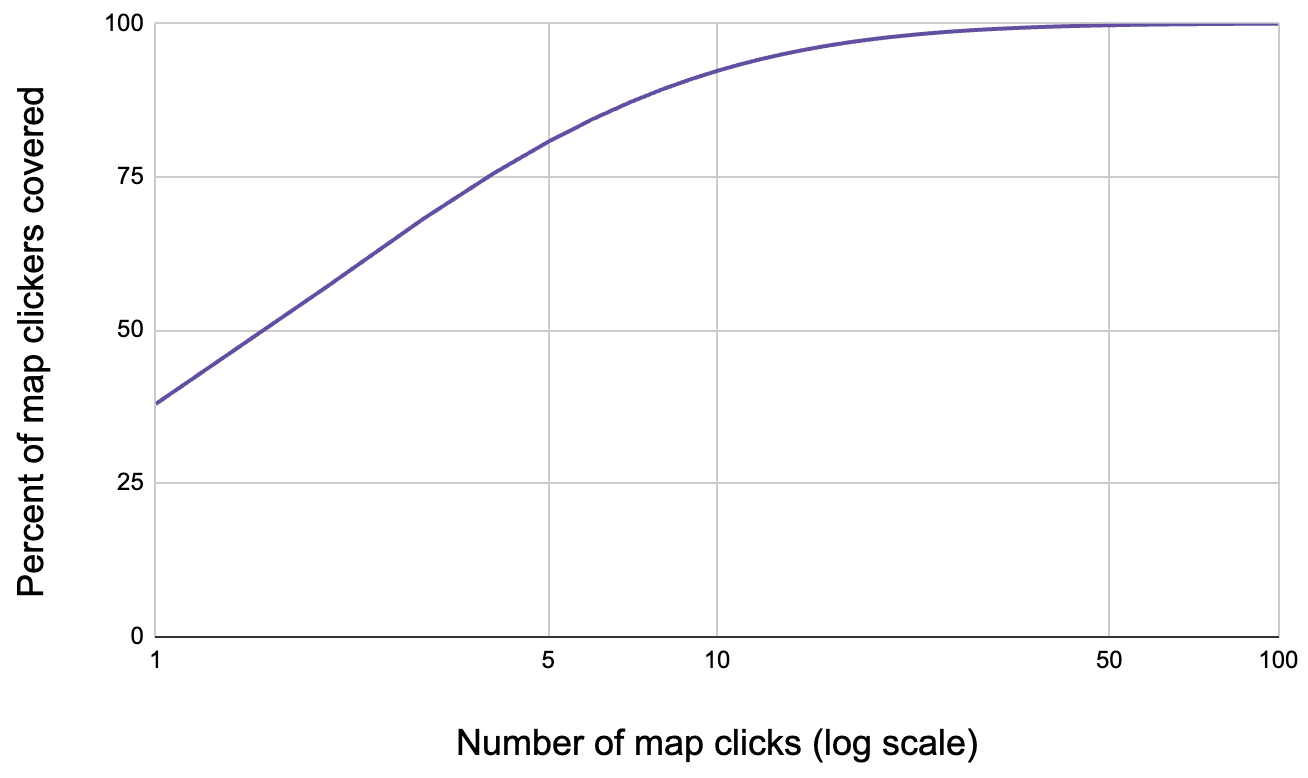}
\caption{\textmd{X-axis: Number of distinct map pins clicks. Y-axis: Percent of map clickers covered. $95\%$ of users click $\le 12$ map pins.}}
\label{fig:nummapclicks}
\end{figure}

Applied by itself, the {\it Less Is More} principle drives down the number of pins towards the degenerate case of a single pin on the map, as that maximizes the average booking probability. So our simplistic view needs a counterbalance which recognizes that users often click through multiple pins. Figure ~\ref{fig:nummapclicks} shows the distribution of the number of distinct pins clicked by searchers. A $95\%$ coverage suggests around $12$ pins. This indicates the optimal user experience requires much fewer than $18$ pins, the fixed number of pins in use for years.

While the arguments above suggest that lowering the number of pins below $18$ might improve user experience, they leave the important question unanswered---that is, how to determine the optimal number of pins for a particular map-result? Intuitively, to construct a map-result we start with the listing that has the highest booking probability, and hence must always be shown on the map. If subsequent listings have comparable booking probabilities, then adding them as map pins provides the user with choice, without degrading the average booking probability too much. However, if booking probabilities of the listings that follow drop significantly, then adding such pins only increase the chance of the user exhausting their attention on such pins, and leaving without booking. We put this intuition into action in Algorithm~\ref{filteralgo}, which we refer to as the {\it Bookability Filter}.

Given two listings $l_x$ and $l_y$, their corresponding ranking model outputs $logit(l_x)$ and $logit(l_y)$ are related to the booking probabilities by Equation~\ref{logiteq}. An in-depth discussion of Equation~\ref{logiteq} can be found in Section 2 of ~\cite{cikmdiversity}.
\begin{equation}\label{logiteq}
logit(l_x) - logit(l_y) = log(P_{booking}(l_x) / P_{booking}(l_y))
\end{equation}
It is convenient to specify the filtering condition in terms of the logits since they are directly available as outputs of the ranking model.

\begin{algorithm}
\SetKwInOut{KwIn}{Input}
\SetKwInOut{KwOut}{Output}
\caption{Bookability Filter}
\label{filteralgo}
\KwIn{A set of $T$ listings $L_{input} = \{l_1, l_2, \dots, l_{T}\}$}
\myinput{Filter parameter $\alpha$}
\KwOut{A set of $N \le T$ listings $L_{output} = \{l_1, l_2, \dots, l_{N}\} $}
$l_{max} \gets \text{argmax}(logit(l_i) : i = 1 .. T)$

\For {$k \gets 1$ until $T$} {
     \If {$logit(l_{max}) - logit(l_k) < \alpha$ } { \label{filtercond}
         $L_{output} \gets L_{output} \cup l_k$
     }
}
\end{algorithm}

For an intuitive understanding of the condition on line~\ref{filtercond} of Algorithm~\ref{filteralgo}, we rewrite it using Equation~\ref{logiteq} as:
\begin{alignat}{2}
\begin{split}
logit(l_{max}) - logit(l_k) & <  \alpha \\
log(P_{booking}(l_{max}) / P_{booking}(l_k)) & <  \alpha \\
P_{booking}(l_{max}) / P_{booking}(l_k) & <  e^{\alpha}  \\
P_{booking}(l_{max}) / e^{\alpha} & < P_{booking}(l_k) \\
\end{split}
\label{pineq}
\end{alignat}

The Bookability Filter admits a listing as a map pin only if its booking probability is greater than $P_{booking}(l_{max}) / e^{\alpha}$. The smaller the value of $\alpha$, the fewer the number of pins on the map; and hence higher their average booking probability, but lower the number of choices available. The $\alpha$ parameter makes the number of map pins dynamic and fine tuned for each map-result, taking into account the booking probabilities of the listings at hand. There is no simple analytical way to infer the optimal value of $\alpha$ that balances both average booking probability and choice. Therefore we shift to determining $\alpha$ empirically using online experiments.

\subsection{Experimental Results}\label{bookabilityresults}
In the first phase, we study the effects of the $\alpha$ parameter through offline simulation of the search system. Table~\ref{tab:alphasweep} summarizes the aggregate statistics of map-results corresponding to different values of $\alpha$. The lower bound of the exploration is set by product experience considerations, and the upper bound the result of diminishing effects. The baseline for all these comparisons is map-results with no filtering and a fixed limit of $18$ pins. 

\begin{table}[!h]
\centering
\begin{threeparttable}
                 \begin{tabular}{lllll}
                   \toprule
                    $\alpha$ & \textbf{$1.0$} & \textbf{$2.0$} & \textbf{$4.0$} & \textbf{$8.0$} \\
                    \midrule
                    Number of map pins            & $-39\%$  & $-25\%$ & $-9\%$ & $-1\%$ \\
                    Average booking probability & $47\%$   & $26\%$ & $8\%$ & $0.7\%$  \\
                    Average total price              & $-19\%$  & $-16\%$ & $-11\%$ & $-3\%$  \\
                    Average number of reviews &  $14\%$   & $10\%$ & $4\%$ & $0.7\%$  \\
                    Average review rating         & $0.05\%$ & $0.09\%$ & $0.13\%$ & $0.03\%$  \\
                   \bottomrule
            \end{tabular}
      \caption{\textmd{Offline exploration of $\alpha$ compared against a baseline with no filtering, which is conceptually equivalent to $\alpha = \infty$.}}
      \label{tab:alphasweep}
\end{threeparttable}
\end{table}

Table~\ref{tab:alphasweep} validates that the Bookability Filter is having the intended effect on map-results. The second phase of our testing investigates how users react to various values of $\alpha$. We run multiple A/B experiments online, where control applies no filtering on map-results, and treatments apply the Bookability Filter with different values of $\alpha$. Table ~\ref{tab:onlinealphasweep} summarizes the effect of $\alpha$ on searchers. A brief explanation of the key metrics evaluated in the online experiments:
\begin{itemize}
\item Uncanceled bookings: This is the top line metric, the number of bookings made by searchers that were not cancelled.
\item 5-star trips: Trips booked by searchers that resulted in 5-star rating after checkout, evaluated $120$ days after end of experiment.
\item Average impressions to discovery: The average number of distinct search results that a booker saw before clicking the listing that was booked. This measures the cognitive load of making a booking.
\item Average clicks to discovery: The number of distinct search results clicked by a booker before clicking the listing that was booked. This is an alternative measure of the cognitive load of making a booking.
\end{itemize}

\begin{table}[!h]
\centering
\begin{threeparttable}
                 \begin{tabular}{llll}
                   \toprule
                    $\alpha$ & \textbf{$1.0$} & \textbf{$2.0$} & \textbf{$4.0$}  \\
                    \midrule
                    Uncanceled bookings            & $1.7\%$ & $1.1\%$ & $0.35\%$ \\
                     5-star trips                             & $1.6\%$ & $1.0\%$ & $0.2\%$ \\
                    Avg impressions to discovery & $-14\%$ & $-9.6\%$ & $-4.0\%$ \\
                    Avg clicks to discovery           & $-5.7\%$ & $-2.5\%$ & $-0.25\%$ \\
                   \bottomrule
            \end{tabular}
      \caption{\textmd{Exploring $\alpha$ through online A/B experiments.}}
      \label{tab:onlinealphasweep}
\end{threeparttable}
\end{table}

In the final phase, we fix the value of $\alpha$ to $1.0$ corresponding to maximum user benefit, and repeat the online A/B experiment at a larger scale, allocating $34$ million searchers worldwide to each of control and treatment. This grinds the p-value of the key metrics below $10^{-5}$. Uncanceled bookings increase by $1.9\%$, measured as a percentage of overall global bookings at Airbnb, making it one of the largest improvements launched over the last several years. 5-star trips increase by $2\%$ indicating not only growth in bookings, but a growth in quality bookings. Average number of results seen by the searcher before clicking on the booked listing reduces by $-16\%$, while search results clicked drop by $-6.8\%$. The reduction in effort to locate the booked listing, due to removal of inferior choices, is the key mechanism driving the gain in bookings.

\section{Making It Robust}
The Bookability Filter puts a lot of faith in the booking probability of the topmost listing, denoted as $P_{booking}(l_{max})$ in Inequality ~\ref{pineq}. We refer to it as the anchor booking probability. At times, the anchor booking probability may be an outlier. An abnormally high anchor booking probability can make it difficult for the rest of the listings to pass through the Bookability Filter, limiting the choice of listings and making searchers quit prematurely.

Gathering insights by debugging a few such cases, we test a more robust way to compute the anchor booking probability---by considering the average of top 3 listings instead of relying on the topmost listing alone. Since median is more resilient to outliers than averages, we further fortify the anchor booking probability to be the median of the top 3 listings. This shifts the anchor booking probability from the topmost listing to the second-highest listing.

\subsection{Experimental Results}
The online A/B test for improving robustness applies filtering with $\alpha$ fixed at 1.0 for both control and treatment map searches. The difference is that in control, the topmost listing supplies the anchor booking probability, while in treatment it is the second listing from the top. This increases the number of map pins in treatment by $6\%$ compared to control, indicating a milder filtering. Searchers to listing viewers conversion improves by $0.15\%$ with a p-value less than $10^{-4}$, with no negative effects on any of the other metrics. The increase in searchers continuing their journey shows that correcting for outliers moderates the cases where filtering was restricting choice for searchers.

\begin{figure}
\includegraphics[height=1.65in, width=2.75in]{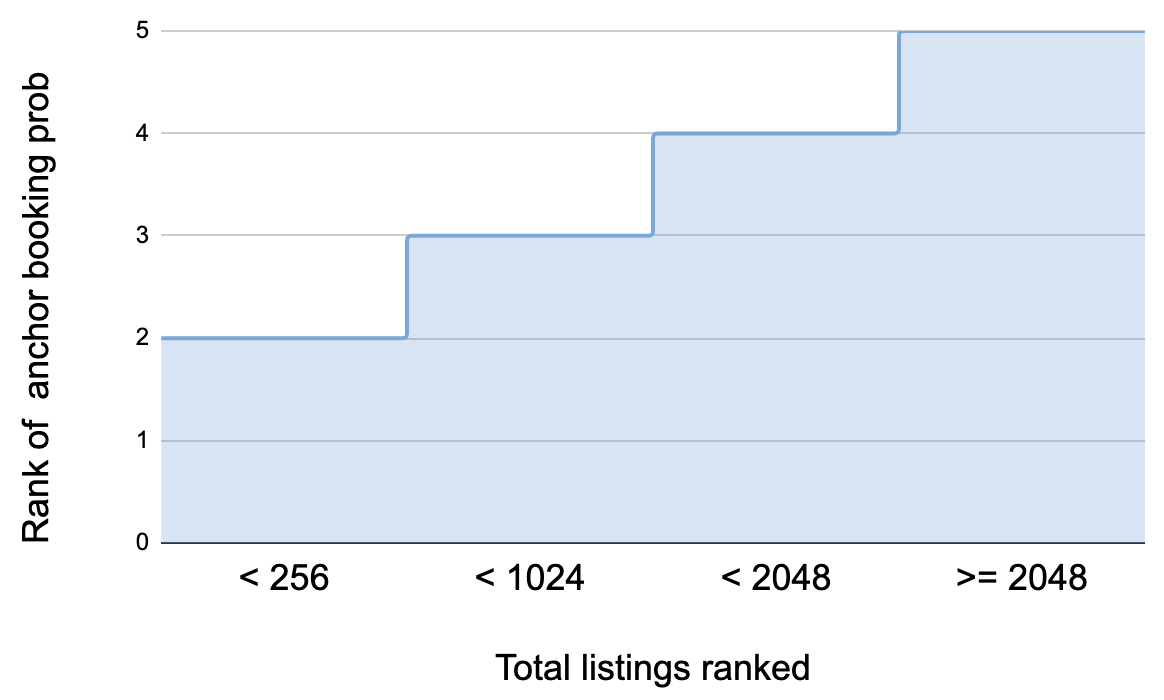}
\caption{\textmd{X-axis: Number of total listings ranked. Y-axis: Rank of the listing supplying the anchor booking probability, tuned through empirical evaluations.}}
\label{fig:adaptivethresh}
\end{figure}

Inspired by the success of this experiment we do one final iteration: instead of fixing the anchor booking probability to the second-highest listing, we make it dependent on the total number of listings in the map area. The reasoning is that the larger the number of listings ranked, more the chances of getting hit by outliers. A plot of the rank of anchor booking probability as a function of the total listings ranked is shown in Figure~\ref{fig:adaptivethresh}. This further increases the number of map pins by $2.6\%$ without degrading any metrics, thus balancing between optimizing average booking probability and providing as many choices as possible.   

\section{Focus vs. Urgency?}
 The blockbuster results of Section~\ref{bookabilityresults} calls for celebration. But they also raise some eyebrows in the room questioning the cause behind such a humongous booking gain. While we can see users clicking less to discover the booked listing, and overall bookings increasing, the metrics don’t communicate what the users are {\it experiencing}.
 
 The optimistic theory is that restricting the map pins is directing the user's attention to the most viable choices, reducing distraction. So users are experiencing {\it focus}. This would mean we are on track for the celebrations.
 
 The alternative theory is that restricting the number of pins is making the users think that not many choices are left for them to book, and what the users are experiencing is {\it urgency}. This would be enough to cancel the party.

The trouble is---how to tell apart the two possibilities?

\subsection{Experimental Results}
We answer the riddle through an online experiment. Let’s walk through an example map-result to understand the experiment design. Suppose there are $1000$ listings available in a map area. After ranking these $1000$ listings by their booking probability, the baseline algorithm picks the top $18$ to show as map pins. Consider the case where the Bookability Filter restricts the number of pins to $14$, leading to a booking gain. The {\it urgency} hypothesis is that reducing the number of pins from $18$ to $14$ makes the user panic, which is the reason behind the booking increase. The {\it focus} hypothesis is that the user has a much higher chance of discovering a bookable listing among the $14$ when compared to sifting through the $18$. The test to differentiate between the two hypotheses is as follows:
\begin{itemize}
\item{$Control$ : Show the 18 map pins.}
\item{$Treatment 1$ : Restrict the map pins to 14, select based on the highest booking probabilities from the baseline 18.}
\item{$Treatment 2$ : Restrict the map pins to 14, matching the number of pins from $Treatment 1$, but select randomly from the baseline 18.}
\end{itemize}
$Treatment 1$ and $Treatment 2$ reduce the number of pins by the same amount, but their average booking probability of map pins is different. There are three possible outcomes of this test, depending on the causes underlying the booking gain:
\begin{itemize}
\item{Urgency \emph{fully responsible} for booking gain: If the booking gain in $Treatment 1$ is completely due to the user experiencing urgency, then the user will experience the same urgency in $Treatment 2$. Both $Treatment 1$ and $Treatment 2$ will show the same booking gain over $Control$ in this scenario.}
\item{Urgency \emph{partially responsible} for booking gain: If urgency provides partial explanation for the booking gain in $Treatment 1$, then a similar gain is expected in $Treatment 2$ as well. We will see a positive booking gain in $Treatment 2$ over $Control$, but less than the gain in $Treatment 1$}.
\item{Urgency \emph{not at all responsible} for booking gain: If the user is not experiencing any urgency in $Treatment 1$, then bookings in $Treatment 2$ will also be devoid of any urgency-based lift, and will be determined by the quality of listings alone. The bookings in $Treatment 2$ will therefore be equal to or less than the bookings in $Control$.}
\end{itemize}

The verdict from the online A/B test is very clear---that urgency is not at all responsible for the booking gain. Bookings for $Treatment 2$ drop by $1.5\%$ compared to $Control$, with p-value below $10^{-6}$. 

\section{Tiered User Attention}\label{minipin}
\begin{figure}
\includegraphics[height=2.5in, width=3.5in]{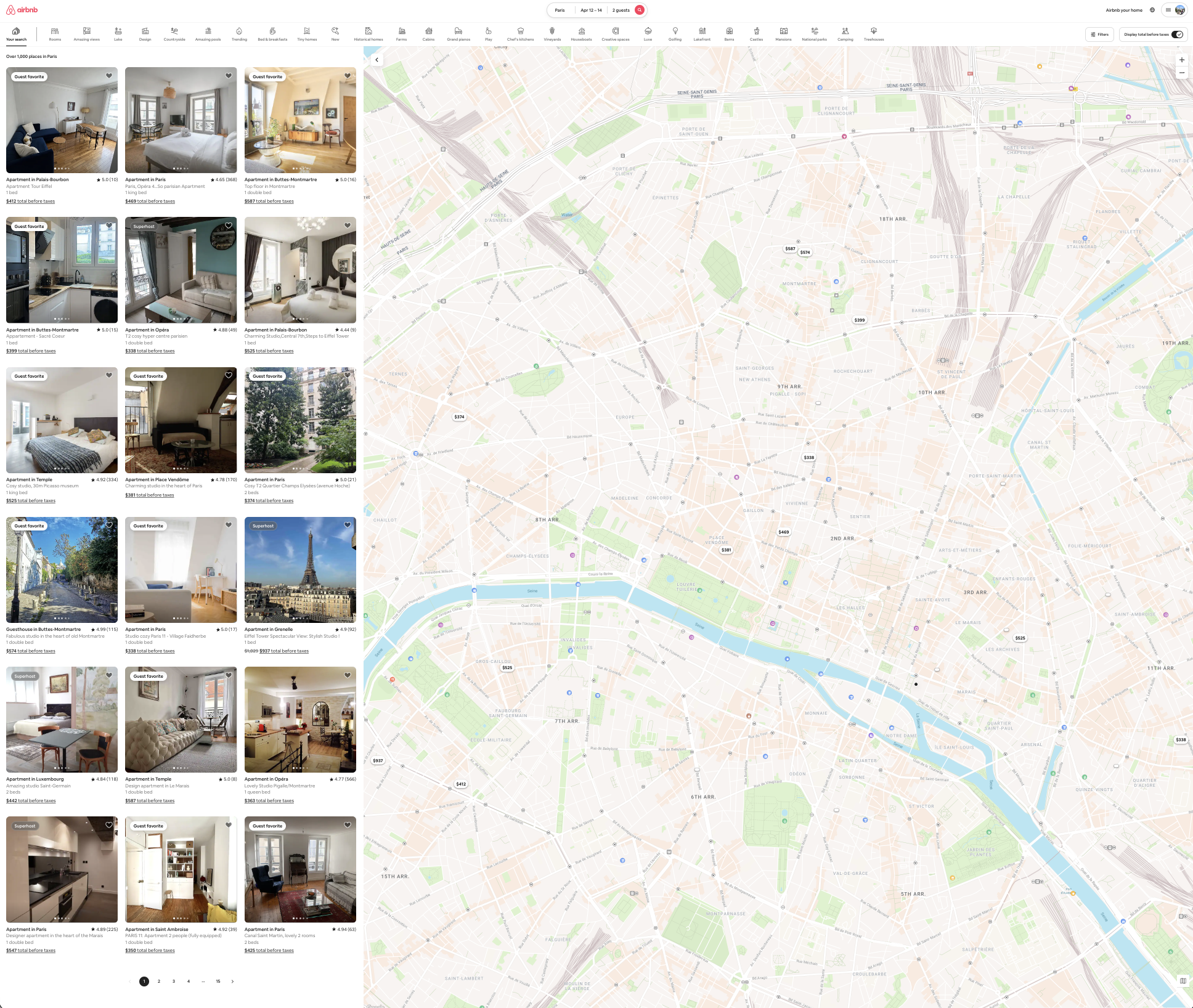}
\caption{\textmd{Simultaneous presentation of list-result and map-result on desktop browsers.}}
\label{fig:desktop}
\end{figure}

When using Airbnb search on web browsers, the results are laid out as a grid of listing cards on the left. On the right the results are displayed on a map (Figure ~\ref{fig:desktop}). Consistency of product experience demands that all the listings shown in the grid, in total $18$, must have a corresponding pin on the map. Lowering the number of map pins is not applicable in this scenario. 

\begin{figure}
\includegraphics[height=1.5in, width=3.0in]{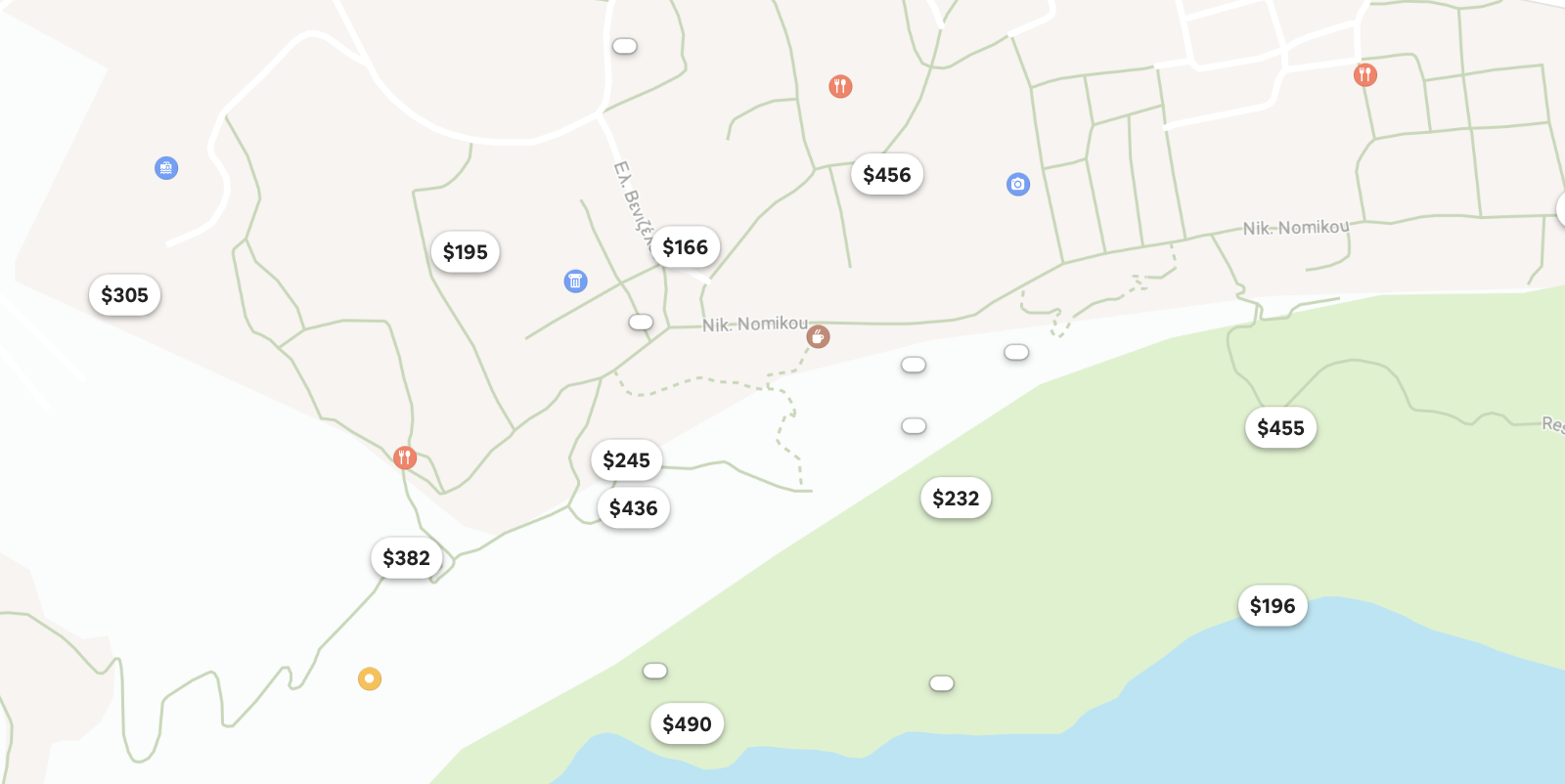}
\caption{\textmd{Map result with mix of regular pins and mini-pins.}}
\label{fig:minipin}
\end{figure}

We adapt the insights from Section~\ref{lim} to the constraints imposed by desktop browsers by creating two tiers of pins: in addition to the ovals with price, we create a smaller oval pin without the price display. We refer to them as mini-pins. Figure~\ref{fig:minipin} depicts a map-result with regular pins and mini-pins. The mini-pins draw less user attention by design; click-through rates for mini-pins is $8$ times less than regular map pins. Differentiating the pins into two tiers allow us to write $P_{attention}(i)$ in Equation~\ref{eq1} as:
\begin{equation}\label{minipineq}
P_{attention}(i) = \begin{cases}
                                1 & \text{if $l_i$ a regular pin} \\
                                1/8 & \text{if $l_i$ a mini-pin} \\
                                0 & \text{otherwise}
                            \end{cases}
\end{equation}

To optimize $P_{booking}(Q)$ in Equation~\ref{eq1}, listings with the highest booking probabilities are assigned regular pins on desktop browsers, while listings with relatively lower booking probabilities are assigned mini-pins. This prioritizes user attention towards the listings with higher chances of getting booked. The $\alpha$ parameter controls the regular pin vs. mini-pin assignment, similar to the Bookability Filter.

\subsection{Experimental Results}
We test the idea through an online A/B experiment where control displays all $18$ listings from the list-result as regular pins on the desktop map. Treatment applies the Bookability Filter to assign listings satisfying Inequality~\ref{pineq} as regular map pins. The remaining listings from the list-result are assigned mini-pins. The results in Table~\ref{tab:minipin} show the effectiveness of tiered user attention.

\begin{table}[!h]
\centering
\begin{threeparttable}
                 \begin{tabular}{ll}
                   \toprule
                    $\alpha$ & \textbf{$1.0$}   \\
                    \midrule
                    Uncanceled bookings            & $0.7\%$   \\
                     5-star trips                             & $0.5\%$   \\
                    Avg impressions to discovery &$ -21\%$   \\
                    Avg clicks to discovery           & $-1.7\%$   \\
                   \bottomrule
            \end{tabular}
      \caption{\textmd{Results from online A/B experiment applying Bookability Filter to differentiate between regular pins vs. mini-pins on desktop.}}
      \label{tab:minipin}
\end{threeparttable}
\end{table}

\section{How Attention Flows On Maps}\label{mapattn}
By distributing the user’s attention over all the map pins in equal measure, Equation~\ref{eq2} led to the {\it Less Is More} principle and generated strong improvements in user experience. Next, we refine our view of how user attention flows across a map by taking inspiration from the rank vs. CTR plots we constructed in Figure~\ref{fig:ctrdiscountcurve}.

\begin{figure}[H]
\centering
\begin{subfigure}{.23\textwidth}
    \centering
    \includegraphics[width=.95\linewidth]{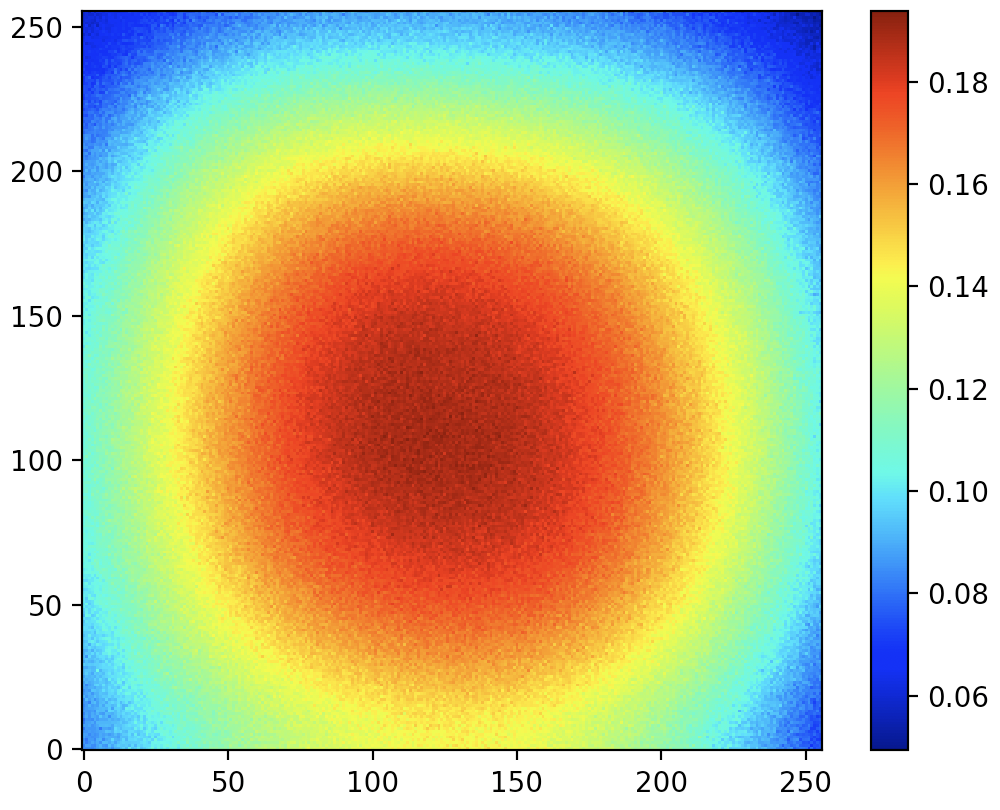}  
    \caption{\textmd{iOS\texttrademark ~app}}
    \label{fig:iosctr}
\end{subfigure}
\begin{subfigure}{.23\textwidth}
    \centering
    \includegraphics[width=.95\linewidth]{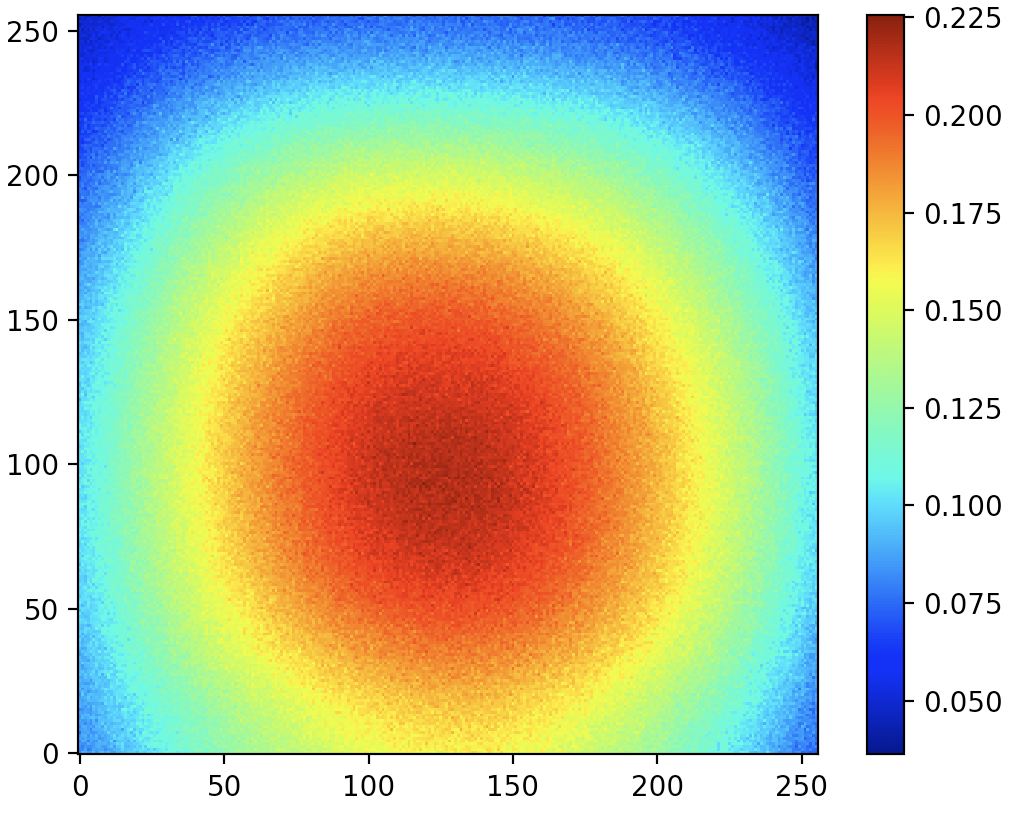}  
    \caption{\textmd{Android\texttrademark ~app}}
    \label{fig:androidctr}
\end{subfigure}
\begin{subfigure}{.23\textwidth}
    \centering
    \includegraphics[width=.95\linewidth]{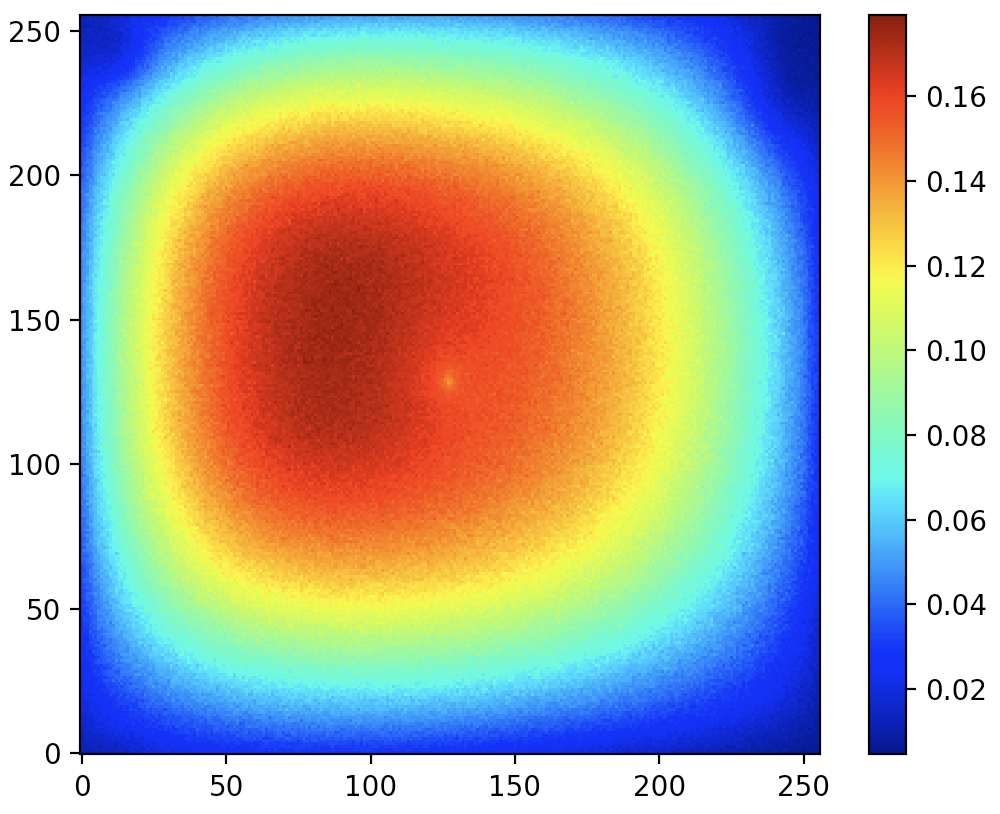}  
    \caption{\textmd{Desktop web browser}}
    \label{fig:desktopctr}
\end{subfigure}
\begin{subfigure}{.23\textwidth}
    \centering
    \includegraphics[width=.95\linewidth]{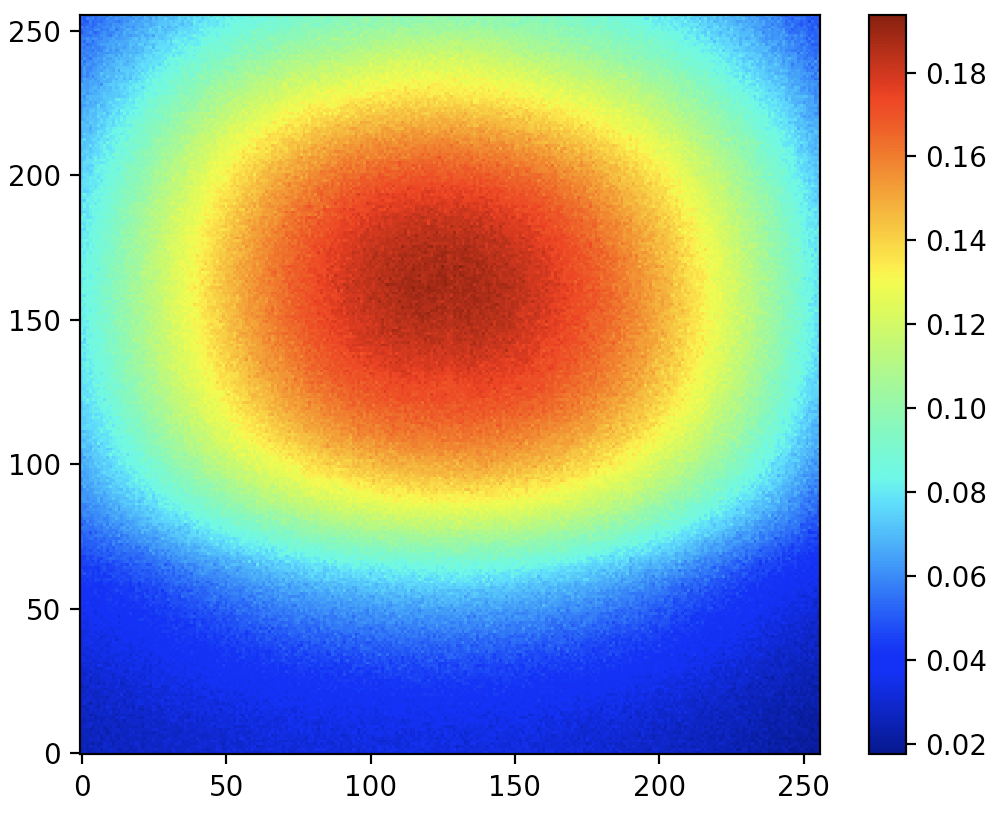}  
    \caption{\textmd{Mobile web browser}}
    \label{fig:mowebctr}
\end{subfigure}
\caption{\textmd{The $(x,y)$ coordinates in each plot corresponds to coordinate on the map, and the value at each coordinate represents the ratio of map pins clicked divided by the number of map pins displayed.}}
\label{fig:2dctr}
\end{figure}

In Figure~\ref{fig:ctrdiscountcurve}, the {\it CTR by position} on the list shows the decay of user attention, going from the top towards the bottom. Explicit eye-tracking based studies of search result pages by others (\cite{papoutsaki2017searchgazer}, ~\cite{lorigo2008eye}) report similar findings. What is the equivalent concept for a map? It is the {\it CTR by coordinates} on the map. This generalizes the concept of attention decay to two dimensions. Figure~\ref{fig:2dctr} shows these 2-D plots of CTR on maps.

The conclusion from Figure~\ref{fig:2dctr} leaps out immediately: user attention is maximum towards the center of the map, and decays radially outwards. Figure~\ref{fig:desktopctr} corresponding to the desktop platform has an additional twist. There is a grid of listings displayed to the left of the map (Figure~\ref{fig:desktop}), which exerts a leftward pull on the user's attention. Figure~\ref{fig:mowebctr} for mobile web browsers exposes an issue: the topmost listing card covers the bottom part of the map, making pins in this region unreachable. The issue is fixed following this discovery.

By aggregating the click behavior of millions of searchers, these plots give the relation between user attention and map coordinates that is agnostic to the map's contents. For any particular individual, landmarks on the map influence the flow of attention as discussed in ~\cite{wenclik2023people}.

\begin{figure}
\includegraphics[height=1.75in, width=3in]{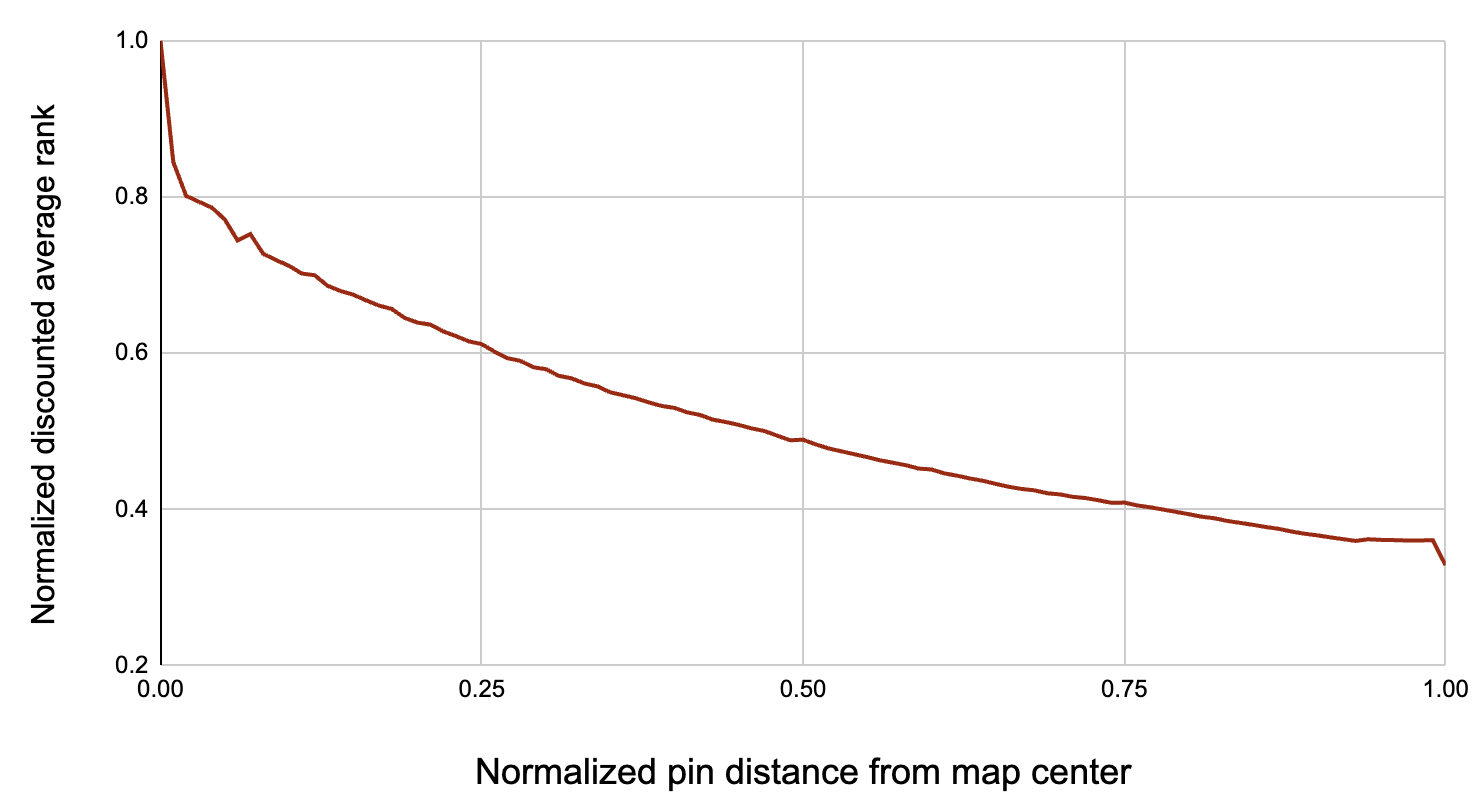}
\caption{\textmd{X-axis: Normalized distance from the map center. Y-axis: For a given distance on the x-axis, we compute the average rank $R_{avg}$ of listings displayed at that distance, then plot $log(2)/log(2+R_{avg})$ on the y-axis. }}
\label{fig:posbydis}
\end{figure}

These plots also solve the mystery posed in Section~\ref{mapnelistexpr}. Click-through rates of map pins drop off with increasing distance from the map center. Listing booking probabilities also go down in a similar manner, as distance of a listing is an important feature contributing to its booking probability (see Figure~\ref{fig:posbydis}). As a consequence, listing rank is correlated with distance from the map center, with higher ranked listings likely to appear closer to the center. This makes the CTR of map pins decay with listing rank, creating the similarity between the map-result and list-result curves in Figure~\ref{fig:mapvslistcurve}. But this similarity is superficial because searchers using the map are completely oblivious to the listing’s rank, as proven by the randomization experiment in Section~\ref{mapnelistexpr}.

\begin{figure}
\includegraphics[height=1.75in, width=3in]{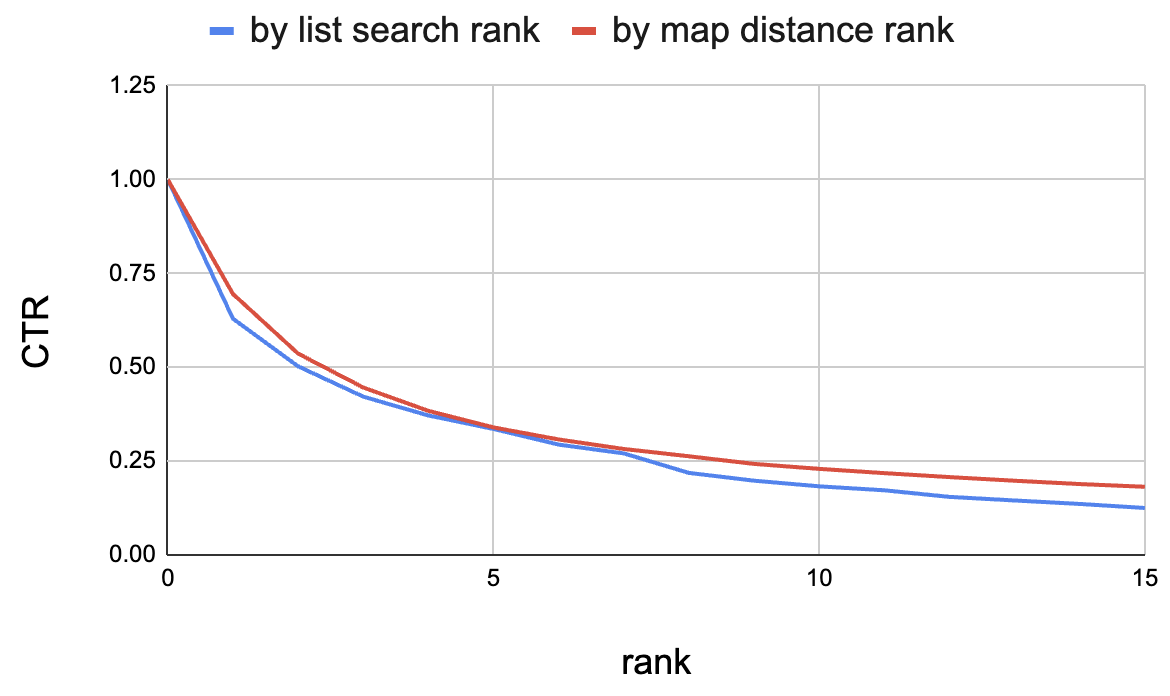}
\caption{\textmd{X-axis: For list-results, the search rank of listings. For map-results, the rank of the map pin by distance from the map center. Y-axis: For list-results the click-through rates of listing cards in search results. For map-results, click-through rates of map pins. }}
\label{fig:ctrbydis}
\end{figure}

Given the two modes of user attention decay, in lists going from top to bottom, and in maps going from center to periphery, the question that arises naturally---how do they compare? But since we plot the decay of user attention in lists as a curve in Figure~\ref{fig:ctrdiscountcurve}, and as a 2-D surface for maps in Figure~\ref{fig:2dctr}, a direct comparison is problematic. To make the comparison, we take map-results and order the map pins by their distance from the map center. This ordering by distance assigns a rank to each map pin, which we use to construct a rank vs. CTR plot for maps. Figure~\ref{fig:ctrbydis} puts the search rank vs. CTR plot for lists next to the distance rank vs. CTR plot for maps.

Search results presented as a list of cards look nothing like pins scattered across a map. Yet, the manner in which user attention decays across the two interfaces in Figure~\ref{fig:ctrbydis} look the same. This suggests that the {\it physical aspect} of attention decay may be tied to the particular interface, for instance 1-D vs. 2-D decay. But the {\it rate} of the attention decay may strongly depend on how much cognitive load users are willing to take to process successive items, a factor independent of the interface. 

Figure~\ref{fig:2dctr} also suggests opportunity to optimize map-results taking into account the decay in user attention. We generalize Equation~\ref{eq1} for map-results as:\\
$\{x(l_i), y(l_i) \} $: Map coordinates for listing $l_i$. \\
$\{ x_0, y_0 \} $: Map center coordinates. \\
$ctr(i, j)$: Click-through rate at offset $\{i, j\}$ w.r.t $\{x_0, y_0\}$. \\
$P_{attention}(l_i)$: Relative attention for $l_i$. \\
\begin{alignat}{2}\label{eq3}
\begin{split}
P_{attention}(l_i) &= \frac{ctr(x(l_i)-x_0, y(l_i)-y_0)}{ ctr(0, 0)} \\
P_{booking}(Q) &= \sum \limits_{i=1}^{N} P_{attention}(l_i) * P_{booking}(l_i)
\end{split}
\end{alignat}

Looking back at Equation~\ref{eq2}, it is a simplified approximation of Equation~\ref{eq3}, although a very effective one in practice. The question now is---how to generate additional improvements based on the insight from Figure~\ref{fig:2dctr}? But optimizing map-results based on Equation~\ref{eq3} turns out to be significantly harder than optimizing list-results based on Equation~\ref{eq1}. In Equation~\ref{eq1}, relative attention only depends on rank $i$ and is independent of the listing. When determining the optimal listing for rank $i$, the ranking algorithm need not worry about altering the user attention associated with rank $i$. In Equation~\ref{eq3}, selecting listing $l_i$ influences both the booking probability $P_{booking}(l_i)$, as well as user attention $P_{attention}(l_i)$, because the user attention is tied to the coordinates of $l_i$. If we have in total $t$ listings eligible for the query $Q$, and have to find the optimal choice of $k$ listings that maximizes $P_{booking}(Q)$ in Equation~\ref{eq3}, then the brute force way would be to evaluate Equation~\ref{eq3} for all ${}^t C_k$ subsets. Given the stringent latency budgets for search, crunching this combinatorial complexity is simply impractical.

\begin{figure}
\centering
\begin{subfigure}{.23\textwidth}
    \centering
    \includegraphics[height=1.6in, width=1.6in]{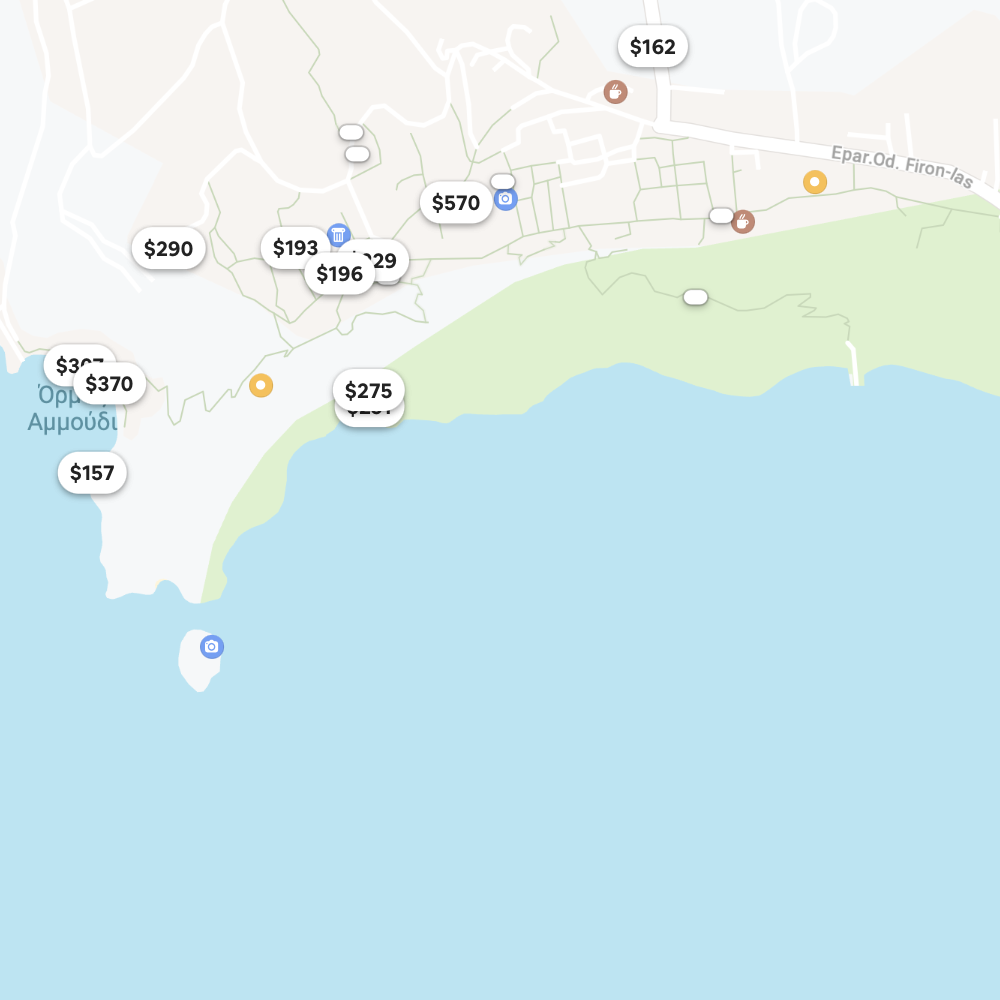}  
    \caption{\textmd{Initial selection.}}
    \label{fig:vp-a}
\end{subfigure}
\begin{subfigure}{.23\textwidth}
    \centering
    \includegraphics[height=1.6in, width=1.6in]{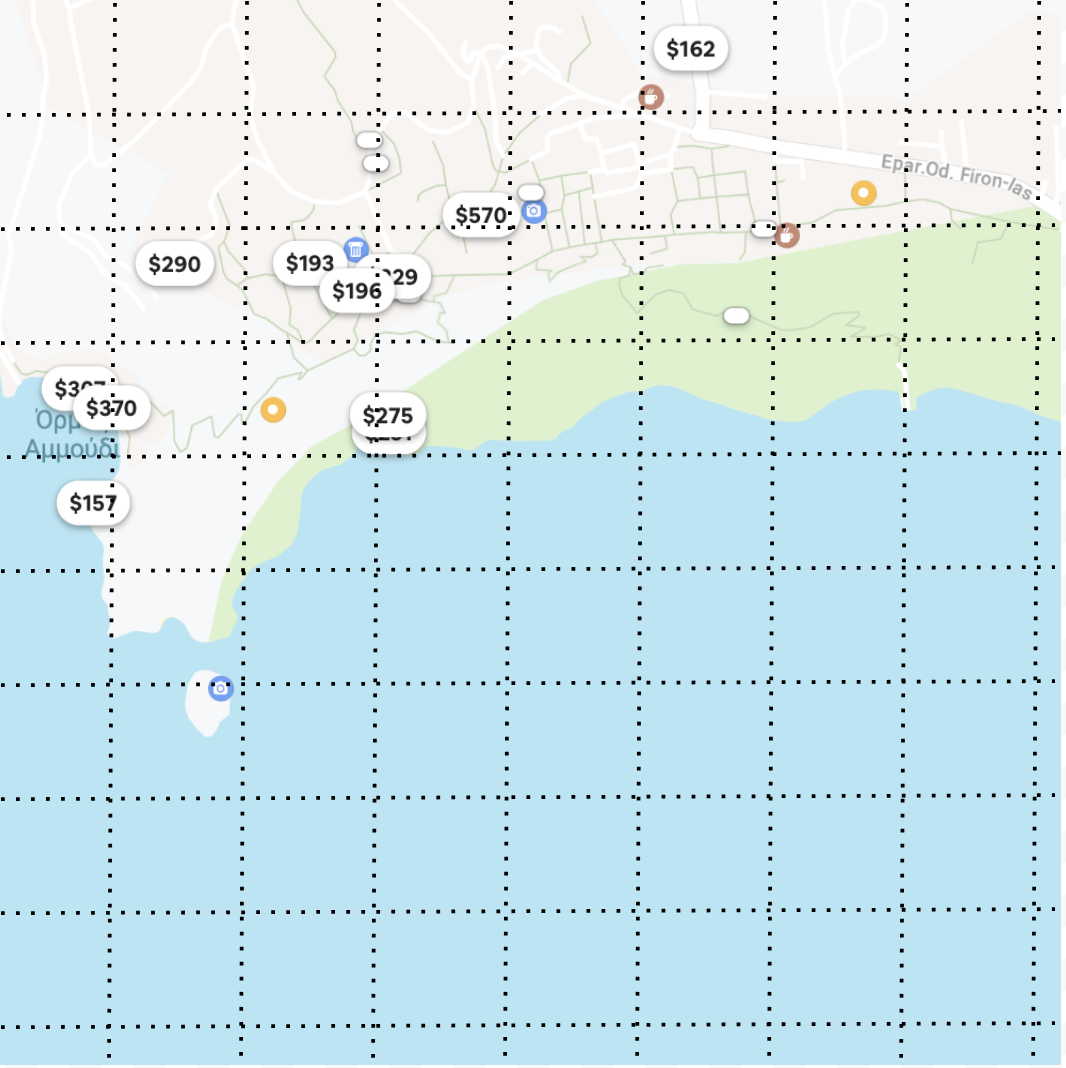}  
    \caption{\textmd{Evaluate potential centers.}}
    \label{fig:vp-b}
\end{subfigure}
\begin{subfigure}{.23\textwidth}
    \centering
    \includegraphics[height=1.6in, width=1.6in]{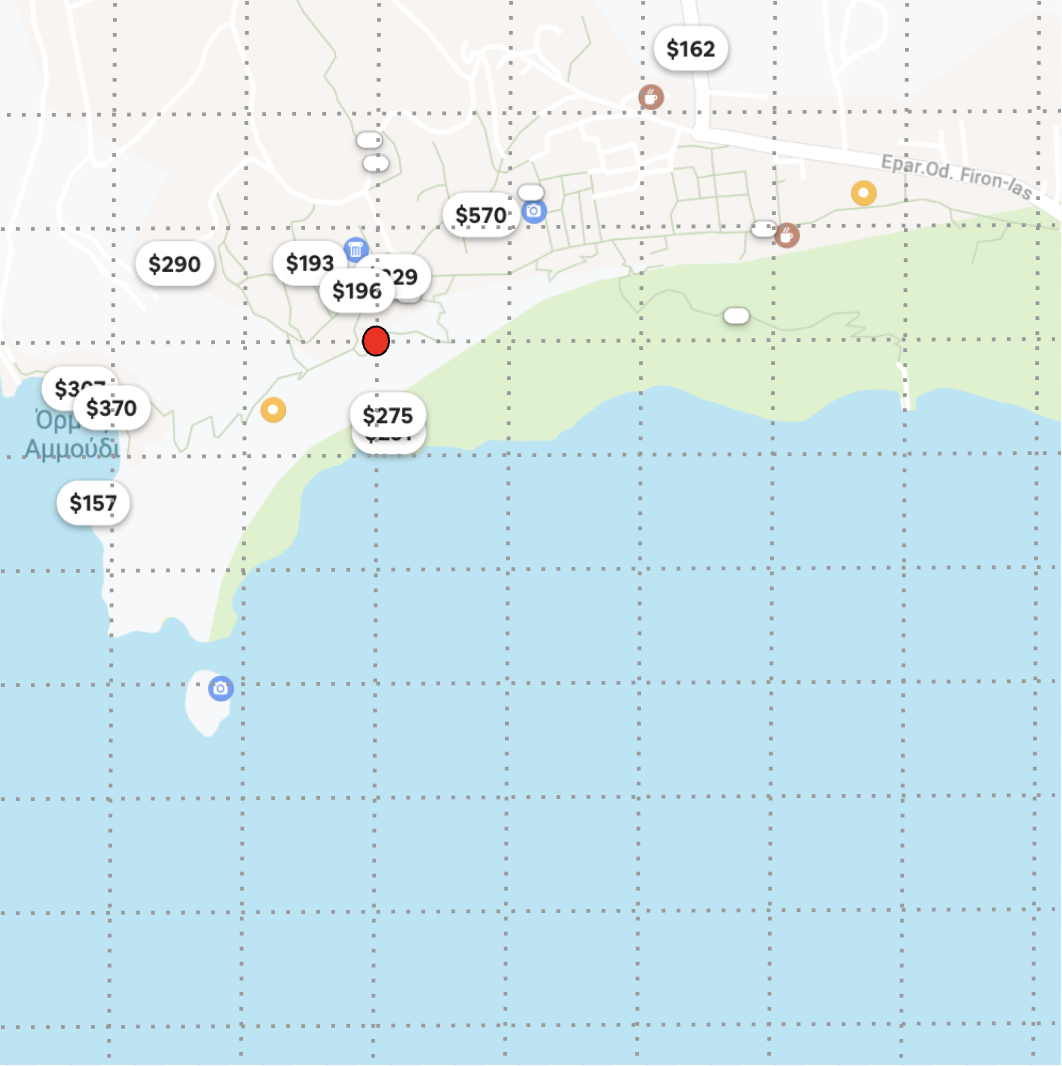}  
    \caption{\textmd{Locate optimal center.}}
    \label{fig:vp-c}
\end{subfigure}
\begin{subfigure}{.23\textwidth}
    \centering
    \includegraphics[height=1.6in, width=1.6in]{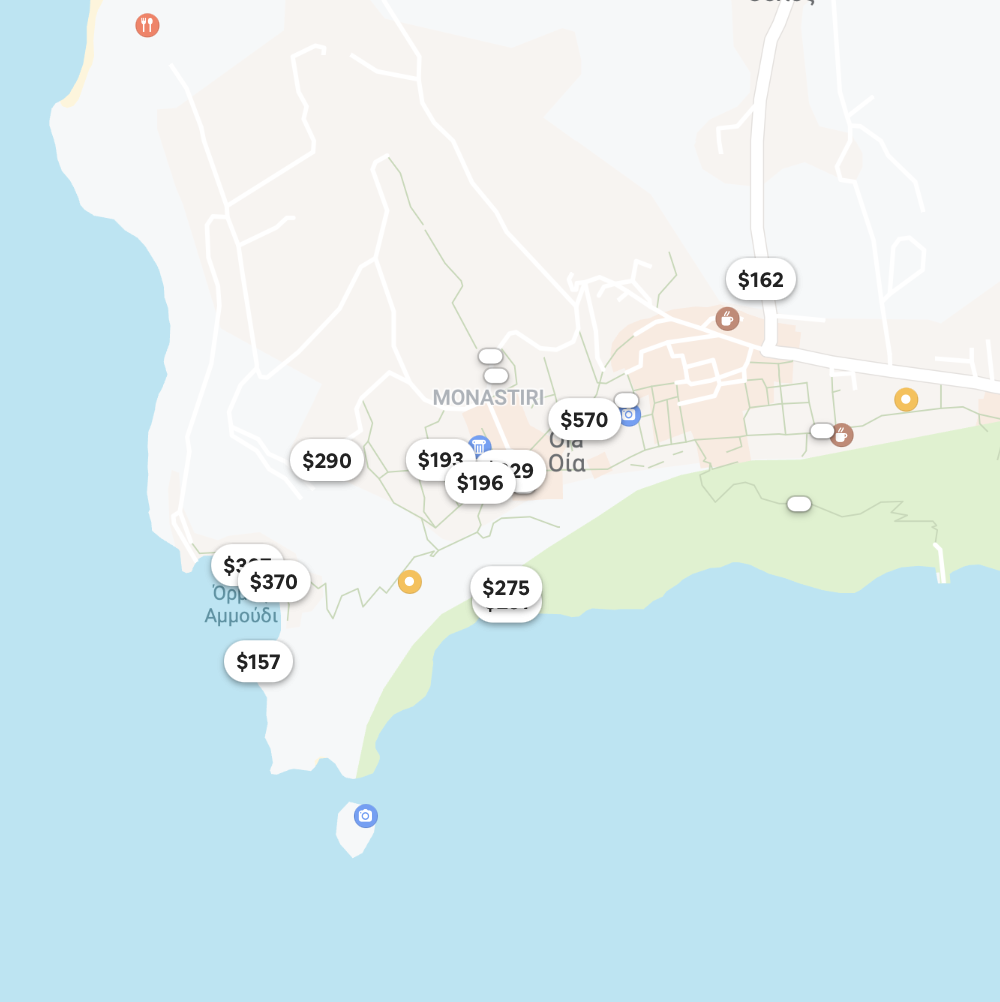}  
    \caption{\textmd{Update center.}}
    \label{fig:vp-d}
\end{subfigure}
\caption{\textmd{An example to illustrate the working of Algorithm~\ref{vpalgo}. (a) Start with an initial bounding box covering the map pins selected based on highest booking probabilities. (b) Iterate evaluate candidates centers over a grid. (c) Locate the center that maximizes $P_{booking}(Q)$ in Equation~\ref{eq3}. (d) Update the bounding box and the map center.}}
\label{fig:algo2example}
\end{figure}

We make the problem tractable by first considering the queries originating from the search box, where we have the freedom to construct the map boundaries. Algorithm~\ref{vpalgo} presents a greedy heuristic as a solution. The heuristic first fixes the choice of listings by selecting them based on booking probabilities alone. It then explores different map centers that optimize for user attention. Figure~\ref{fig:algo2example} gives a pictorial view of Algorithm~\ref{vpalgo}. The heuristic simplifies the problem by optimizing $P_{booking}(l_i)$ and $P_{attention}(l_i)$ in a disjoint manner. Overall complexity of the heuristic is $O(k^2n)$, where $k$ is the map width divided by the $\epsilon$ parameter, and $n$ is the number of output map pins. 

\begin{algorithm}
\SetKwInOut{KwIn}{Input}
\SetKwInOut{KwOut}{Output}
\caption{Greedy heuristic for map-result optimization}
\label{vpalgo}
\KwIn{A set of $t$ listings $L_{input} = \{l_1, l_2, \dots, l_{t}\}$}
\myinput{Iteration step size $\epsilon$}
\KwOut{A set of $n \le t$ listings $L_{map} = \{l_1, l_2, \dots, l_{n}\} $}
\myoutput{Map center $\{x_0, y_0\}$}
$L_{sorted} \gets SortBy(P_{booking}(l_i), l_i \in L_{input}) $

$L_{map} \gets \{l_1, l_2, \dots, l_{n}, l_i \in L_{sorted}\}$

$x_{min}, x_{max} \gets min(x(l_i)), max(x(l_i)), l_i \in L_{map}$

$y_{min}, y_{max} \gets min(y(l_i)), max(y(l_i)), l_i \in L_{map}$

$maxBooking, x_0, y_0 \gets - \infty $

\For {$i \gets x_{min}$ ; $i < x_{max}$ ; $i  \gets i +  \epsilon $} {

    \For {$j \gets y_{min}$ ; $j < y_{max}$ ; $j  \gets j +  \epsilon $} {
         $P_{attn}(l_k) \gets \frac{ctr(x(l_k)-i, y(l_k)-j)}{ ctr(0, 0)}, l_k \in L_{map} $
         
         $bkNew \gets \sum P_{attn}(l_k) * P_{booking}(l_k), l_k \in L_{map} $ 
    
         \If {$maxBooking < bkNew$ } { 
               
               $x_{0} \gets i$
               
               $y_0 \gets j$
               
               $maxBooking \gets bkNew$
               
         }
     }
 }
\end{algorithm}

Implicit search queries generated by the user while pointing the map to an area are simpler to optimize, as they have fewer degrees of freedom. The map area and the position of the candidate map pins are fixed in such cases. Including a ranking model feature for the distance of a pin from the map center is able to factor in the decay of attention in Figure~\ref{fig:2dctr}, since the booking label already contains the information. To study this aspect, we compare ranking models with and without the feature in the next section.

\subsection{Experimental Results}
For queries generated by the user's map movements, the control of the online A/B experiment is a ranking model that does not have the distance of a pin from the map center as a feature. Treatment is a ranking model including that feature. Uncanceled bookings increase by $0.27\%$ in treatment, along with an increase of $2\%$ in click-through rates of map pins.

For queries originating from the search box, we do an online A/B experiment with Algorithm~\ref{filteralgo} as the control and the treatment applies Algorithm~\ref{vpalgo} to further optimize the map. We observe an increase of $0.39\%$ in uncanceled bookings for new guests with a p-value of 0.006. Users new to Airbnb find the placement of the most bookable listings at the map center particularly helpful as their map moves decrease by $1.5\%$, and their chance of encountering areas with very few listings decrease by $0.8\%$.

The results indicate that the techniques discussed in Section~\ref{lim} and Section~\ref{minipin} provide the biggest opportunities for optimization, with the techniques discussed in Section ~\ref{mapattn} providing further incremental gains.

\section{Insights And Future Work}
Models of user attention developed for lists do not apply for map interfaces. This opens up a new research area where user attention on maps can be modeled as:
\begin{itemize}
\item Distributed {\it uniformly:} Summarized as Equation~\ref{eq2}, it leads to Property~\ref{prop1} and the Bookability Filter in Section~\ref{lim}.
\item Distributed {\it discretely:} Applicable for differentiated map pins that are useful when outright filtering is not desirable. Captured by Equation~\ref{minipineq} in Section~\ref{minipin}.  
\item Distributed {\it continuously:} Based on Figure~\ref{fig:2dctr} in Section~\ref{mapattn}, Equation~\ref{eq3} forms the basis for Algorithm~\ref{vpalgo}. 
\end{itemize}
Equation~\ref{eq1} is the foundation tying all these insights together. The literature on improving ranking mostly focuses on the $P_{booking}(l_i)$ term in Equation~\ref{eq1}, whereas this paper refines the $P_{attention}(i)$ term for maps. This multi-year track of research generated some of the most impactful launches in the history of search ranking at Airbnb.

And yet, this is merely a start that points to several exciting research potentials for the future. To name a few, significant work has gone into making learning to rank unbiased for list-results (\cite{ltrdebias}, ~\cite{ltrdebias2}, ~\cite{ltrdebias3}). A similar need exists for map-results. Section~\ref{minipin} proved the effectiveness of separating the map pins into two tiers. This suggests scope to further demarcate the map pins to direct the user attention towards more relevant ones. For map-results, the relative ordering of the top results is not as consequential. It is possible that loss functions that take advantage of this aspect and optimize for top-k precision (\cite{berrada2018smooth}, ~\cite{garcin2022stochastic}) outperform the straightforward pairwise cross-entropy loss. In Section~\ref{mapattn} we described a heuristic that was able to improve upon the baseline, raising hopes for other heuristics that might generate further improvements.

Given this is a completely new area, the lack of established public benchmarks is a challenge. Creating a dataset to help research in this area is on our roadmap.

%\end{document}  % This is where a 'short' article might terminate